\documentclass[aps,prl,showpacs,twocolumn,floats]{revtex4}
\usepackage{graphicx}
\usepackage{amsmath}
\usepackage{psfrag}

\newcommand{\be}{\begin{equation}}              
\newcommand{\ee}[1]{\label{#1} \end{equation}}  

\newcommand{\bee}{\begin{eqnarray}}             
\newcommand{\eee}{\end{eqnarray}}               
\def\reff#1{(\ref{#1})}

\begin{document}

\title{Temporal chaos versus spatial mixing in reaction-advection-diffusion systems}
\author{Arthur V. Straube\footnote{E-mail: straube@stat.physik.uni-potsdam.de \\[0.5mm] {\mbox{} Paper published in Phys. Rev. Lett {\bf{93}},
174501 (2004)}}, Markus Abel and Arkady~Pikovsky}
\affiliation{Department of Physics, University of Potsdam,
 Am Neuen Palais 10, PF 601553, D-14415, Potsdam, Germany}
\begin{abstract}
We develop a theory describing the transition to a spatially homogeneous regime
in a mixing flow with a chaotic in time reaction. The
transverse Lyapunov exponent governing the stability of the homogeneous state
can be represented as a combination of Lyapunov exponents for spatial mixing
and temporal chaos. This representation, being exact for time-independent flows
and equal P\'eclet numbers of different components, is demonstrated to work
accurately for time-dependent flows and different P\'eclet numbers.
\end{abstract}

\maketitle

Complex spatiotemporal dynamics has attracted large interest in
the last decades. Recently, reaction-diffusion equations have been
subject of intense research due to their rich variety of patterns.
They describe many important physical systems, such as chemical
reactions~\cite{Kapral-Showalter-95}, lasers~\cite{Arecchi91}, or
semiconductors~\cite{Scholl-01}. On the other hand, complex
structures can be created in fluid mechanics by spatial
mixing~\cite{Ottino-89}. In this Letter, we consider the combined
action of the above effects, leading to a
reaction-advection-diffusion system, see Eq.~\reff{radsys} below.
Such systems have been investigated with respect to front
propagation~\cite{Abel-Celani-Vergni-Vulpiani-01}, excitable
dynamics~\cite{Neufeld-etal-02}, and the filamental structure of
reactive particles \cite{Toroczkai-etal-98}. Practically, stirred
flows with reaction are relevant from large scales (plankton
dynamics in the oceans~\cite{Abraham-98}) to microscales
(construction of a lab on a chip~\cite{Knight-eyal-98}), and are
important for biophysical, ecological, and chemical applications;
similar equations describe the dynamo effect in
magnetohydrodynamics~\cite{Childress-Gilbert-95}.

In this Letter, we consider a temporally chaotic reaction process.
It is known that in presence of diffusion, temporal chaos can lead
to the appearance of nontrivial spatial structures and space-time
chaos. We demonstrate that such structures can appear in the
presence of mixing, too. We develop a theory for the transition
from spatially homogeneous (fully mixed) temporally chaotic state
to a nonhomogeneous one, and compare it with calculations. Our
approach is strongly related to the theory of complete
synchronization of coupled chaotic systems, which is significantly
extended because the spatial mixing leads to special types of
coupling.

We now formulate the reaction equations. The evolution of the
concentrations $\phi_i$, $i=1,\ldots,N$ due to reaction is described by a nonlinear system %
\be
\frac{d{\phi}_i}{dt} = F_i(\phi_1,\ldots,\phi_N)\;,
\ee{reac}
with regular or chaotic solutions $\phi^0_i(t)$. Additionally,
each component is subject to diffusion and advection by an
incompressible velocity field $\mathbf{v}(\mathbf{r},t)$.
Normalization by the characteristic advection time allows a
description by dimensionless diffusion constants $d_i$ (equivalent
to P\'eclet numbers $d_i^{-1}\sim{\rm Pe}_i$, generally different)
and the dimensionless reaction rate, or Damk\"ohler number, Da.
The resulting spatio-temporal equations are
\be {\partial \phi_i\over \partial
t}+(\mathbf{v}\cdot\nabla)\phi_i= d_i\nabla^2\phi_i +{\rm Da\,}
F_i(\phi_1,\ldots ,\phi_N)\;. \ee{radsys}
We assume that the concentrations do not influence the flow, so
that the field $\mathbf{v}(\mathbf{r},t)$ does not depend on
$\phi_i$. We supply Eq.~\reff{radsys} with no-flux boundary
conditions $\nabla\phi_i|_S=0$ on the boundary $S$ (periodic
boundary conditions are  straightforwardly treated in an analogous
way).

For a homogeneous spatial distribution of concentrations $\phi_i$
the advection and diffusion term vanish and the equations
\reff{radsys} reduce to \reff{reac} with rescaled time.  To study
the stability of spatially homogeneous solutions $\phi_i^0({\rm
Da}\,t)$ of \reff{radsys} we linearize the equations near this
solution and obtain for a small perturbation field
$\varphi_i(\mathbf{r},t)$
\be {\partial \varphi_i\over \partial t}+
(\mathbf{v}\cdot\nabla)\varphi_i=d_i\nabla^2\varphi_i +{\rm
Da\,}\, C_{ij}({\rm
  Da} \,t)\varphi_j\;,
\ee{lin}
where $C_{ij}(t)=\frac{\partial F_i}{\partial \phi_j}$ is the
Jacobi matrix of the system \reff{reac} on the solution
$[\phi_1^0(t),\ldots,\phi_N^0(t)]$. Generally, the solutions of
\reff{lin} grow or decay exponentially in time $||\varphi||\sim
e^{\lambda t}$, where $\lambda$ belongs to the spectrum of
Lyapunov exponents (LE). Clearly, the LEs of the solution of
\reff{reac} belong to this spectrum, describing growth or decay of
homogeneous perturbations. The stability of the homogeneous
solution towards inhomogeneous perturbations is described by the
largest LE corresponding to a spatially varying Lyapunov vector
$\varphi(\mathbf{r},t)$ -- the transverse LE $\lambda_\perp$ [in
the sense that it is transverse to a manifold of spatially
homogeneous solutions of \reff{lin}]. For diffusion constants and
time-dependent flow $\mathbf{v}(\mathbf{r},t)$ given, the
transverse LE can be determined only numerically. For certain
situations, considered below, we obtain this exponent
analytically.

We start the analysis of \reff{lin} with the simplest case, a
time-independent velocity field
$\mathbf{v}=\mathbf{v}(\mathbf{r})$ and equal diffusion constants
$d_i=d$. In this case the ansatz
$\varphi_i(\mathbf{r},t)=X(\mathbf{r})\Phi_i(t)$ allows for a
separation of time and space  dependence of the perturbation
field. The spatial component is determined by the
advection--diffusion eigenvalue problem
\be d\nabla^2X-(\mathbf{v}\cdot\nabla)X=-\Gamma X,\qquad \nabla
X|_S=0\;. \ee{evpx}
The eigenvalue $\Gamma$ describes the decay of non-homogeneous
states of the passive scalar field $u(\mathbf{r},t)$, governed
by\\[-6mm]
\be {\partial u\over\partial
t}+(\mathbf{v}\cdot\nabla)u=d\nabla^2u\;, \ee{ps}
because for an exponentially time-dependent solution, \reff{ps}
reduces to \reff{evpx}. This problem has been recently analyzed
in~\cite{Pikovsky-Popovych-03}.  The eigenvalue $\Gamma=0$
corresponds to the spatially homogeneous solution and does not
contribute to the stability of inhomogeneous perturbations. For
the latter, the mode corresponding to the smallest non-zero
eigenvalue of \reff{evpx} is relevant, we denote it by $\gamma$.
As has been argued in~\cite{Pikovsky-Popovych-03}, this eigenvalue
crucially depends on the nature of flow $\mathbf{v}(\mathbf{r})$,
and thus on Pe. If the flow is mixing, the eigenvalue $\gamma$ is
well separated from zero even for small diffusion constant $d$.
However, a flow typically contains chaotic and regular domains
(islands of Kolmogorov-Arnold-Moser tori). Then, for small
diffusion constants, there are weakly decaying modes concentrated
in these islands, so that $\gamma$ may be rather small. We notice,
that in terms of effective  diffusion (coarse grained on the
system size $L$) this eigenvalue can be represented as
$\gamma=d_{eff}L^{-2}$~\cite{biferale-crisanti-vergassoloa-vulpiani-95}.

The equation for the temporal part $\Phi_i$ is\\[-5mm]
\be {d\Phi_i\over dt}=-\gamma\Phi_i+{\rm Da\,} C_{ij}({\rm
Da}\,t)\Phi_j \ee{tp1} \\[-5mm]
With the ansatz $\Phi_i=e^{-\gamma t} w_i$ this equation is
transformed to the equation for linear perturbations of the
reaction problem \reff{reac}\\[-5mm]
\be {d w_i\over dt}= {\rm Da\,} C_{ij}({\rm Da\,}t) w_j\;,
\ee{odelin}\\[-4mm]
its asymptotic solution is $||w|| \sim e^{{\rm Da}\lambda t}$, where
$\lambda$ is the largest LE of the attractor in \reff{reac}.  Thus,
for the perturbation we have $\Phi_i\sim e^{({\rm Da}\,\lambda-\gamma)
t}$ and the explicit formula for the transverse LE reads:
\be
\lambda_\perp={\rm Da\,}\cdot\lambda-\gamma({\rm Pe})\;.
\ee{trle1}
The stability condition of the spatially homogeneous state can be
formulated as $\lambda_\perp<0$. If the oscillations of the
concentrations are regular, then the largest Lyapunov exponent is
nonpositive $\lambda\leq 0$ and this regime is always stable
against spatially inhomogeneous perturbations. A nontrivial
transition occurs for a chaotic regime, if $\lambda>0$. Here the
stability condition leads to the critical value\\[-6mm]
\be
{\rm Da}_{cr}= \frac{\gamma({\rm Pe})}{\lambda}\;.
\ee{rcr}
A similar condition for a trivial case of reaction-diffusion
system has been obtained in \cite{Pikovsky-84b} and for an
abstract mapping model of mixing in \cite{Pikovsky-92a}. We
emphasize that condition \reff{rcr}, obtained for a realistic
reaction-advection-diffusion system, can be directly applied to an
experiment. Indeed, the eigenvalue $\gamma$ can be directly
measured from the time evolution of the contrast of a passive
scalar in the flow under investigation
\cite{Rothstein-Henry-Gollub-99}. The LE $\lambda$ can be
determined from the advection-free setup: the critical domain size
$L_c$ at which the patterns appear is related to $\lambda$ via
$C\,d\,L_c^{-2}=\lambda$, where $C$ is a geometrical factor
depending on the domain form.


The analysis above is based on simplifying assumptions:
time-independence of the velocity field and equality of diffusion
constants.  In general, if the velocities are time-dependent and
the diffusion constants are different, Eq.~\reff{lin} cannot be
simplified and should be analyzed numerically. Since we are
interested in stability with respect to spatially inhomogeneous
perturbations, the solution should be sought in the class of
fields $\varphi$ having zero spatial average (the spatially
homogeneous modes in \reff{lin} are decoupled from other modes).
Numerically, one can use the usual method for calculation of the
largest LE: starting with an arbitrary initial field in
\reff{lin}, with vanishing spatial average, one integrates
\reff{lin} along with \reff{radsys} performing normalization of
the linear field to avoid numerical over- or underflow; averaging
the logarithm of the normalization factors yields the transversal
LE $\lambda_\perp$.

We apply this numerical method to the time-dependent flow,
$2\pi$-periodic in space, suggested in
\cite{Antonsen-Fan-Ott-Garcia-Lopes-96}:
\be \mathbf{v}=\left\{a_x f(t)\cos[y+\theta_x(t)],\;a_y
[1-f(t)]\cos[x+\theta_y(t)]\right\}. \ee{flow}
Here, the simple analytical analysis above is not applicable.
Depending on the functions $f(t),\theta_x(t),\theta_y(t)$, the
flow can be time periodic or irregular. In the former case, the
particle trajectories demonstrate typical Hamiltonian dynamics
with chaotic regions and stability islands
(see~\cite{Antonsen-Fan-Ott-Garcia-Lopes-96,
Pikovsky-Popovych-03}). A weakly turbulent irregular flow occurs
if the functions $f(t),\theta_x(t),\theta_y(t)$ are random
functions of time. We have considered both of these cases with the
reaction dynamics \reff{reac} given by the Lorenz equations.

\begin{figure}[t]
\centerline{\includegraphics[width=0.32\textwidth]%
{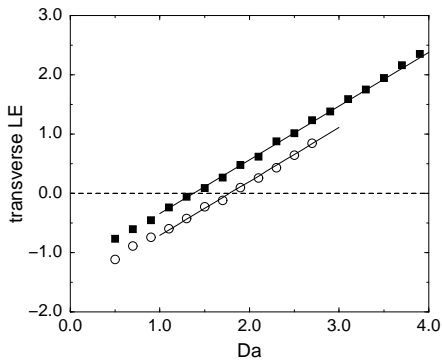}} \caption{Numerical calculation of transverse LEs
from Eqs. ~\reff{lin} and from approximations \reff{trle1} and
\reff{trle2}. Parameters: $a_x=a_y=20$, $\langle
T_{int}\rangle=0.5$. Squares: $d_i=0.1$; circles: $d_1=0.1$,
$d_2=0.2$, $d_3=0.5$. Lines are according to \reff{trle1} and
\reff{trle2}.}
 \label{fig:leperp}
\end{figure}

An irregular flow was modeled by setting $f(t)$  as a
$(0,1)$-telegraph process with independent exponentially
distributed time intervals $T_{int}$, and independent uniformly
distributed phases $\theta_x, \theta_y \in (0,2\pi)$. The
transverse Lyapunov exponent has been calculated as described
above; the results are shown in Fig.~\ref{fig:leperp}. Remarkably,
$\lambda_\perp$ is nearly a linear function of Da like in
\reff{trle1}. Moreover, we can demonstrate numerically that the
formula \reff{trle1} is valid also quantitatively when all the
diffusion constants are equal. To this end, we have calculated the
decay rate $\gamma$ from the linear evolution Eq.~\reff{ps} of the
passive scalar. Now this equation does not yield the simple
eigenvalue problem \reff{evpx} because the velocity field is time
dependent, but defines in full analogy with the discussion above
the asymptotic decay rate $\gamma$ in the sense of the LE
(cf.~\cite{Pierrehumbert-94}):
\be
\gamma=-\lim_{t\to\infty}\frac{\ln||u(\mathbf{r},t)||}{t}\;.
\ee{ledef}
Thus, $\gamma$ is physically interpreted as the asymptotic decay
rate of the contrast of the passive scalar in the
advection-diffusion problem \reff{ps}. In Fig.~\ref{fig:leperp},
we compare the transverse LE $\lambda_\perp$, calculated from
\reff{trle1}, with the  numerical estimation of $\lambda_\perp$.
Here, the value of $\gamma$ has been calculated according to
\reff{ledef}, $\lambda$ is the largest LE of the Lorenz model; the
correspondence with the numerics is within statistical errors.
This result indicates that the time dependencies due to chaotic
time evolution of species and due to irregularity of the flow are
essentially ``separable.''

The analysis above must be modified if the diffusion constants in
\reff{lin} are different. In this case one has different passive scalar
evolution problems Eq.~\reff{ps}, and Eq.~\reff{tp1} is
modified to
\be
{d\Phi_i\over dt}=-\gamma_i\Phi_i+{\rm Da}\,C_{ij}({\rm Da}\,t)\Phi_j\;.
\ee{tp2}
The generalized transverse LE for~\reff{tp2} is
\be \Lambda_\perp({\rm Da},\gamma_1,\ldots,\gamma_N)\;, \ee{lgen}
for $\gamma_1=\ldots=\gamma_N=\gamma$ it reduces to \reff{trle1},
but generally cannot be related to the LE of the reaction system
like in \reff{odelin}. As a result, instead of (\ref{trle1}) and
(\ref{rcr}), we obtain the following condition for the critical
Damk\"ohler number \be \Lambda_\perp({\rm
Da}_{cr},\gamma_1,\ldots,\gamma_N)=0\;. \ee{lgen1}

There is a connection of the defined above transverse LEs with the theory
of synchronization. In the latter one considers two coupled nonlinear systems
\be
\begin{array}{l}
\displaystyle
\frac{d{\phi}_i^{(1)}}{dt} = F_i(\phi_j^{(1)})+
\frac{\gamma_i}{2}(\phi^{(2)}_i-\phi^{(1)}_i)\;,\\[0.5em]
\displaystyle
\frac{d{\phi}_i^{(2)}}{dt} = F_i(\phi_j^{(2)})+
\frac{\gamma_i}{2}(\phi^{(1)}_i-\phi^{(2)}_i)\;,
\end{array}
\ee{reac2}
and looks for the stability of the completely synchronized state
$\phi_i^{(1)}(t)=\phi_i^{(2)}(t)$. Then the linearized equation for perturbations
coincides with  \reff{tp2}. Thus, the generalized transverse LE $\Lambda_\perp$
determines the synchronization threshold. This analogy shows that in a stirred
reaction the role of coupling constants $\gamma_i$ is played by effective decay
rates where both advection and diffusion contribute.

In general, some coupling constants can be absent, i.e. the
corresponding coefficients $\gamma_i$ vanish. In the context of
chaotic mixing, such a situation appears, e.g., for surface
reactions.  Here, it can happen that mobility of some chemical
species is large so that they are advected by the fluid flow,
while other species are so strongly chemisorb that they cannot
move laterally across the surface (see,
e.g.,~\cite{Hildebrand-Mikhailov-Ertl-98}).  These chemisorbed
species are described by vanishing spatial decay $\gamma_i$.  In
the biological context, mixed and nonmixed components interact,
e.g., in marine sediments by tidal flows.

\begin{figure}[ht]
\centerline{\includegraphics[width=0.26\textwidth]%
{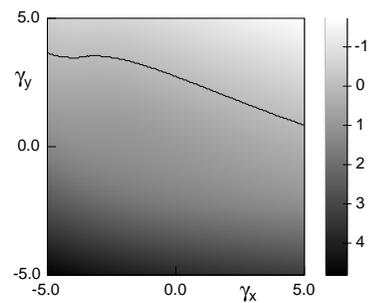}} \caption{Generalized transverse LEs for the Lorenz
model according to \reff{tp2} with $\gamma_z=0$ and ${\rm Da}=1$.
The zero contour level is shown with a line.}
 \label{fig:gtrle}
\end{figure}

Notice, that although a transformation
$\Phi_i=e^{-s t} W_i$ does not allow us to reduce \reff{tp2} to \reff{odelin}, one
parameter, say $\gamma_N$, can be eliminated with such a transform. As a result,
the generalized transverse LE \reff{lgen} can be represented as a function of   $N-1$
parameters only:
\be
\Lambda_\perp({\rm Da},\gamma_i)=
{\rm Da}\Lambda_\perp\left(1,\frac{\gamma_1-\gamma_N}{{\rm Da}},
\ldots,\frac{\gamma_{N-1}-\gamma_N}{{\rm Da}},0\right)-\gamma_N\;.
\ee{trle2}
In Fig.~\ref{fig:gtrle}, we show $\Lambda_{\perp}$ for the Lorenz
system. The above result for $\Lambda_{\perp}$ agrees well with
the direct calculation of transverse LEs for the full system
\reff{lin}, see Fig.~\ref{fig:leperp}.

\begin{figure}[ht]
\centerline{\includegraphics[width=0.33\textwidth]%
{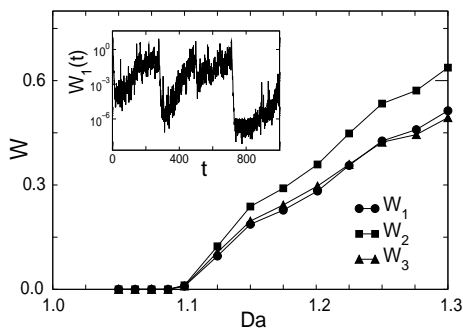}} \caption{Time-averaged contrast of the reaction
species vs the Damk\"ohler number [the critical value is is 1.096,
cf. \reff{rcr}]. The inset shows temporal intermittency of the
contrast $W_1$ for ${\rm Da}=1.15$. Parameters: $d_i=0.04\cdot
\pi^2$, $a_x=a_y=1.2\cdot \pi$.}
 \label{fig:contrast}
\end{figure}

Above we have performed the stability analysis of the spatially
homogeneous regime. Next we discuss properties of structures that
appear beyond the stability threshold. A natural quantity to
characterize the inhomogeneity of the pattern is the contrast, or
variance, $W_i=\langle (\phi_i-\langle\phi_i\rangle)^2\rangle$. We
solved \reff{radsys} using \reff{flow} with the periodic function
$f(t)=(0,1)$ like in \cite{Antonsen-Fan-Ott-Garcia-Lopes-96} with
unit period and constant $\theta_{x,y}$, and the Lorenz model as
reaction. All three components show the same onset, consistent
with linear theory, see Fig.~\ref{fig:contrast}. Near the onset of
spatial inhomogeneity, the temporal behavior of the pattern
contrast is highly intermittent (see inset in
Fig.~\ref{fig:contrast}). As in a transition to complete
synchronization, the reason for this intermittency is the
fluctuations of the local exponents $\Lambda_\perp$ and
$\gamma_i$, similar to previously investigated
cases~\cite{Pikovsky-84b, Fujisaka-Yamada-87}.

We have demonstrated that the transition to spatially
inhomogeneous structures in a mixed flow with chaotic in time
evolution of concentrations is determined by the transverse LE.
This exponent can be with good accuracy represented by a sum of
``temporal'' and ``spatial'' contributions according to
\reff{trle1}, or, more generally, to \reff{trle2}. We can give the
following physical arguments in favor of this effective
``decoupling'' allowing for the separation ansatz. (i) For large
P{\'e}clet numbers, although different species have different
diffusion constants, their mixing properties are to a large extent
determined by advection. Thus, one can expect that spatial
structures of Lyapunov vectors for different variables
$\varphi_i(\mathbf{r},t)$ in \reff{lin} are close to each other.
This is confirmed by numerical simulations: we have calculated the
spatial correlation coefficients between different variables for
the example presented in Fig.~\ref{fig:leperp} (the case of
different diffusion constants) and have found for these
correlations the values larger than $0.83$. The same property can
be expected for small P{\'e}clet numbers, where the spatial
structure is given by the purely diffusive version of \reff{evpx}
and one has the same modes for different species, only the decay
constants are different. (ii) Although stirring for a
time-dependent flow is not constant in time, this seems to have a
very small effect on the average transverse LE. We have checked
this by solving the generalized system \reff{tp2} where the decay
rates $\gamma_i$ are not constants but oscillating functions of
time. In some range of periods, the corresponding LE
$\Lambda_\perp$ remained the same within $1\%$ even if the
modulation was as large as $100\%$.

Concluding, we have applied the method of generalized transverse
Lyapunov exponents to the analysis of transition from homogeneous
to inhomogeneous field in a stirred, temporally chaotic chemical
reaction. Our theory is valid for general chaotic reactions and
general mixing setups -- the latter do not need to be perfect
(hyperbolic). For equal diffusion constants, the critical value of
the Damk\"ohler number \reff{rcr} has a clear physical meaning: it
determines, whether the growth rate due to temporal chaos (given
by the maximal LE of the nonlinear system) wins over the decay
rate due to mixing (given by the decay rate of the contrast of a
passive scalar in the flow). For different diffusion constants, or
for surface reactions where only some species are stirred, the
critical Damk\"ohler number can be formulated as a novel problem
in the theory of complete synchronization, whose solution is
expressed in terms of generalized transverse LEs.

We acknowledge fruitful discussions with B.~Eckhardt, O.~Popovych,
T.~Tel, and G.~Zaslavsky. A.~S. thanks DAAD for support. A.~S. and
M.~A. were supported by the German Science Foundation.


\end{document}